\documentclass[aps,pre,twocolumn,groupedaddress,floatfix,superscriptaddress]{revtex4}
\pdfoutput=1

\usepackage{dcolumn}
\usepackage{amsmath}
\usepackage{amssymb}
\usepackage{lipsum}

\usepackage{graphicx}
\usepackage{bm}
\usepackage[T1]{fontenc}
\usepackage{color}
\usepackage{url}
\usepackage[bf]{subfigure}
\usepackage{rotating}
\usepackage{soul}
\usepackage{scalefnt}
\usepackage[hidelinks ]{hyperref}
\usepackage{scalefnt}
\usepackage{tabularx}

\usepackage{perpage}
\MakePerPage{footnote}

\usepackage{color}
\definecolor{redcolor}{rgb}{1.0,0.,0.}

\newcommand{\sTHD}{\sigma_{\text{\tiny THD}}}
\newcommand{\uTHD}{\mu_{\text{\tiny THD}}}

\begin{document}
\title{A Pattern Recognition Approach to Transistor Array Parameter Variance}
\author{Luciano da F. Costa}
\affiliation{S\~ao Carlos Institute of Physics, University of S\~ao Paulo, S\~ao Carlos, SP, Brazil}

\author{Filipi N. Silva}
\affiliation{S\~ao Carlos Institute of Physics, University of S\~ao Paulo, S\~ao Carlos, SP, Brazil}
\author{Cesar H. Comin}
\affiliation{S\~ao Carlos Institute of Physics, University of S\~ao Paulo, S\~ao Carlos, SP, Brazil}

\begin{abstract}
The properties of bipolar junction transistors (BJTs) are known to vary in terms of their parameters.  In this work, an experimental approach, including pattern recognition concepts and methods such as principal component analysis (PCA) and linear discriminant analysis (LDA), was used to experimentally investigate the variation among BJTs belonging to integrated circuits known as transistor arrays.  It was shown that a good deal of the devices variance can be captured using only two PCA axes.  It was also verified that, though substantially small variation of parameters is observed for BJT from the same array, larger variation arises between BJTs from distinct arrays, suggesting the consideration of device characteristics in more critical analog designs.  As a consequence of its supervised nature, LDA was able to provide a substantial separation of the BJT into clusters, corresponding to each transistor array.  In addition, the LDA mapping into two dimensions revealed a clear relationship between the considered measurements.  Interestingly, a specific mapping suggested by the PCA, involving the total harmonic distortion variation expressed in terms of the average voltage gain, yielded an even better separation between the transistor array clusters.
\end{abstract}

\maketitle

\setcounter{secnumdepth}{1}

\section{Introduction}

Bipolar transistors are at the root of modern electronics, but the variability of their parameters remains an important phenomenon with important theoretical and practical implications.  Perhaps as a consequence of the adoption of negative feedback in promoting circuit invariance to transistor parameters, the variability of transistor behavior has received relatively little attention in the literature more recently.  There have been several works addressing this problem at a larger integration and processing scales (e.g.~\cite{jaeger1997microelectronic}), typically considering digital applications, but relatively fewer works have addressed discrete device variance in analog electronics.  Yet, a better understanding of transistor variability at the individual level remains an important issue because it can help, in a number of ways, the design of better electronic circuits (e.g.~\cite{chen2016active}).  Ultimately, all large scale integration systems are composed by a massive number of individual transistors.  For instance, the knowledge of the current gain of an individual (or a lot of) BJT provides valuable subsidies for deciding how much gain can be sacrificed to negative feedback in exchange for circuit improvements.  In addition, recent findings~\cite{costa2016negative} suggest that the latter technique may have limited effects in ensuring transistor parameter invariance, motivating further related research, as well as design procedures that take into account the specific characteristic curves and parameters of each individual device. 

Interestingly, though important issues such as transistor parameter dependency have been addressed in semiconductor physics, it usually does not go to the level of probing industrialized devices, so there is substantial space for research aiming at bridging this gap.  Fortunately, many advances in instrumentation~\cite{bell1994electronic}, signal analysis techniques~\cite{oppenheim2010discrete,proakis2006digital}, powerful multivariate statistical~\cite{joliffe1992principal,johnson2002applied} and pattern recognition~\cite{bishop2006pattern,witten2005data} methods have paved the way to devising experimental approaches that can contribute to our understanding of industrialized transistor variability.    In a recent study~\cite{costa2016negative}, we showed that discrete BJTs variance is relatively constrained within each transistor type, but much larger among different types of transistors.  These differentiating variabilities are such that well-defined clusters of transistors could be identified, each of such clusters corresponding to respective transistor types.  However, the intra-cluster variance resulted markedly distinct for different types, resulting in clusters of different sizes and shapes.  One of the important consequence of such findings, at least for the considered devices and configurations, is that transistor variability is not as large as to prevent statistical methods to be applied to characterize specific types.  It was also shown that the variability of BJTs is mostly a two-dimensional phenomenon, in the sense that two linear combinations of traditional transistor characteristics was found to be enough to account for 88\% of the variability of the considered BJTs.  Despite such interesting findings, an important question remained unanswered mainly concerning the possible causes of BJT variability.  More specifically, it would be interesting to investigate to what an extent such a variance depends on transistors sharing or not the same substrate.  After all, devices sharing the same substrate have been exposed to almost identical fabrication conditions.   

Here, we resort to pattern recognition concepts and methods, including principal component analysis (PCA) and linear discriminant analysis (LDA), to investigate \emph{transistor arrays}, namely integrated circuits containing several BJTs in a common substrate, therefore providing an interesting laboratory to study transistor variability.  Such types of ICs were produced during the 80's and 90's for several applications, such as buffering, driving, power control, audio, among other possibilities.   One of the advantages of such arrays, as indicated in the respective datasheets, would be more controlled device characteristics as a consequence of the common substrate and fabrication conditions shared by the devices.  However, to our knowledge, few (or no) systematic experimental data is available in the literature regarding the characteristics of transistor arrays. 

So, transistor arrays constitute an interesting resource to complement our previous investigations of transistor variability.  More specifically, here we experimentally obtain characteristic curves for 6 devices in 50 transistor arrays, totaling 300 BJTs.  Voltage and current gains, as well as Total Harmonic Distortion (THD) measurements were obtained respectively to a modified common emitter class A configuration.  Several features of each transistor were then numerically estimated, and several modern pattern recognition concepts and methods --- including PCA and LDA --- were used to investigate clustering and variability of individual transistors.  Several interesting results are reported.  First, we have that, as expected, the parameter variability among transistors in a same array is much smaller (by an approximate factor of 20 with respect to current gain) compared to the overall variability among transistors from all arrays.  The overall distribution of parameters was analyzed using PCA and subsequent density estimation, and found to be possibly distributed between two main peaks, suggesting the existence of two prototypes, at least for the considered BJT samples and conditions.  The variation among the arrays was found not to be negligible, indicating that more critical analog projects involving more than one array may justify acquiring and taking into account the experimental characteristics of the used devices.  Further analysis by using LDA revealed an improved seggregation between the transistor arrays, as well as a clear relationship between the measurements.  An even better separation between the transistor arrays was achieved by mapping the THD in terms of a combination of measurements derived from the PCA.  The obtained results indicate that a good deal of transistor parameter variation may stem from different substrates, probably as a consequence of residual doping variations or other industrial processing events. 

The current article is organized as follows.   First, we present how the experimental data was obtained and processed in order to estimate the respective electrical features.  Then, we show the results obtained by using PCA and LDA, leading to the conclusion that the BJTs in each transistor array are indeed much more similar one another than devices devices from distinct ICs.

\section{Materials and Methods}

\begin{figure}
  \centering    \includegraphics[width=0.35\textwidth]{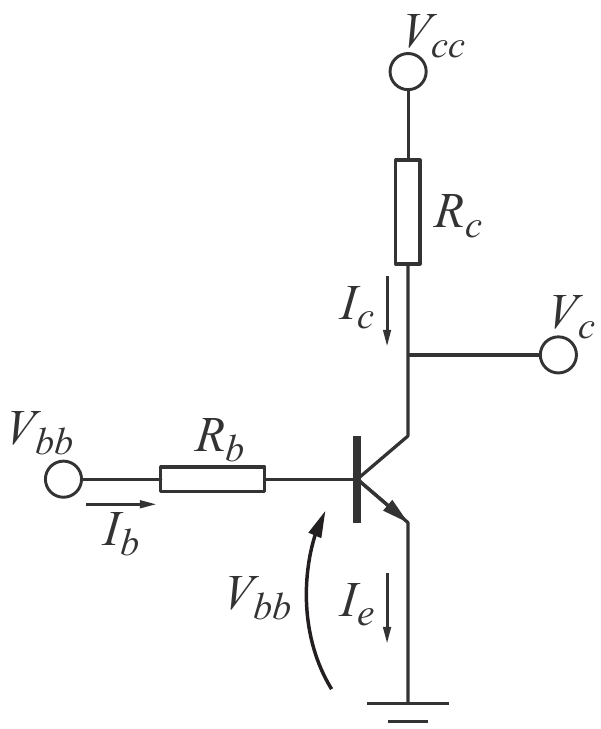}
  \caption{The modified, fixed bias, class A common emitter configuration adopted in the reported experiments.}\label{f:circuit}
\end{figure}

A customized stimuli generator and data acquisition was designed and implemented for the purpose of the reported experiments.  It consists of a modular 16-bit microprocessor system, with separated digital and analog parts.  All supplies are thoroughly decoupled.  The analog inputs are sample-and-holded and buffered into 4 channels, while the analog outputs are buffered by voltage followers.  A precision reference source is used for DA conversion, taking place over 4 independent channels.  Acquired data is buffered to ensure real-time sampling, and then transferred to SD cards.  The characteristics curves are obtained by subsequent fixed $V_{cc}$ scans.  A modified version of the fixed bias, class A common emitter configuration~\cite{boylestad2002electronic} (shown in Figure~\ref{f:circuit}) is used for such purposes, using $R_b = 11900 \Omega$ and $R_c = 671 \Omega$.  By \emph{modified} we mean the use of the base resistor $R_b$ as a current limiter to the input signal, used fixed instead of emitter biasing. 

\begin{figure*}
  \centering    \includegraphics[width=0.85\textwidth]{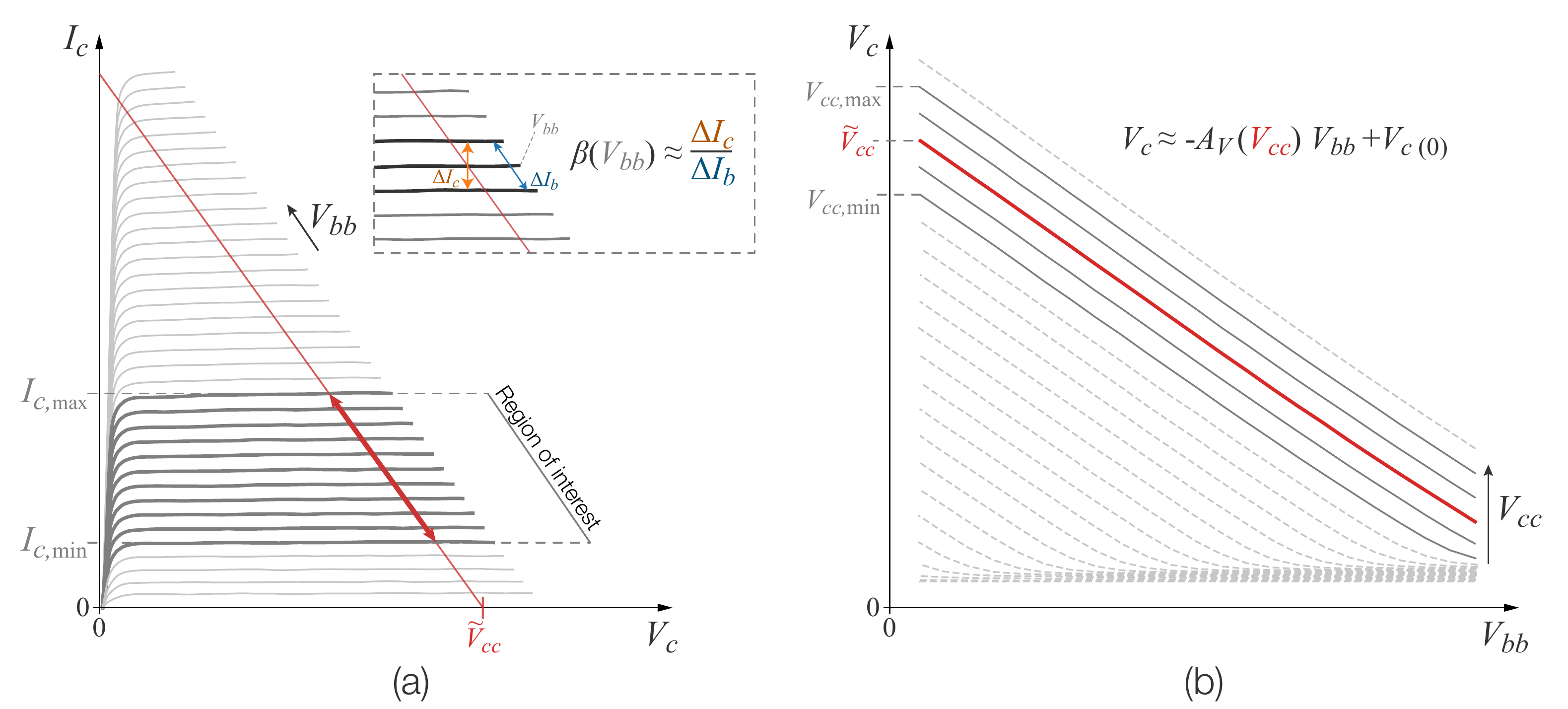}
  \caption{A typical set of isolines and definition of the region where our measurements are taken.  The inset illustrates the estimation of the current gain $\beta$.}\label{f:methods}
\end{figure*}

In order to estimate the properties of the considered transistors (please refer to Figure~\ref{f:circuit}), we start by organizing their characteristic curves in terms of isolines with fixed $V_{bb}$. Next, we select a representative contiguous set of isolines defined by a range in $I_c$. This is done so that the influence from the saturation region over the transistor parameters is reduced. Figure~\ref{f:methods}(a) illustrates a typical set of isolines obtained for BJT transistors. The figure also shows a possible choice for the region of interest corresponding to isolines far away from the saturation region as well as from the cutoff.  The reported experiments adopted $I_{c,min} = 2mA$,  $I_{c,max} = 4mA$ and  $\tilde{V}_{cc} = 6V$.

The total current gain $\beta$, defined as $dI_c/dI_b$, can be estimated directly from the obtained isolines. The inset of Figure~\ref{f:methods} illustrates the procedure to obtain $\beta$. For a chosen $V_{cc}$ far from the saturation region, and for given reference $V_{bb}$ isoline, we estimate $\beta(V_{bb})$ as
\begin{equation}
\beta(V_{bb}) \approx {\Delta I_c \over \Delta I_b} \label{eq:Beta}
\end{equation}
where $\Delta I_c$ and $\Delta I_b$ corresponds respectively to the changes of $I_c$ and $I_b$ between two isolines: one sequentially before and another after the reference isoline.  Such a procedure yields a distribution of $\beta$ along $V_{bb}$ from which we can obtain statistics such as the average, $\mu_{\beta}$, and standard deviation, $\sigma_{\beta}$.

The voltage gain $A$ is the ratio between the input voltage and the resulting output voltage, i.e., $A = dV_c/dV_{bb}$.  Observe that such an adopted voltage gain differs from the more commonly used $A_v$ derived from emitter biased configuration, consider the base voltage directly instead of the voltage at the incoming lead of $R_b$ in the adopted configuration shown in Figure~\ref{f:circuit}. Such a property can be estimated by means of the transfer functions $V_c(V_b)$. Such transfer functions can be retrieved from the characteristic curves by taking the values of $V_b$ along a load line, which in turn are defined by $V_cc$ and $R_c$. More specifically, $A(V_{cc})$ can be estimated as the negative of the slope of each transfer curve corresponding to values of $V_{cc}$. Figure~\ref{f:methods}(b) illustrates typical transfer functions obtained for BJTs. Minimum least squares approximation can be used to fit the transfer curves to straight lines and recover the slope. Similarly to $\beta$, we also considered a distribution of the values for the voltage gain, however in this case, by varying $V_{cc}$. Therefore, we can define the average, $\mu_{A}$, and standard deviation, $\sigma_{A}$, of the voltage gain. 

Another important characteristic usually expected from transistor and respective circuits is linearity.  This property can be quantified by the so-called total harmonic distortion (THD)~\cite{cordell2011designing}. This property is calculated as follows. A pure sinusoidal function with frequency $f$ is applied to the system. Then, the amplitudes of the fundamental ($V_f$) as well as of the harmonics generated on the output ($V_{2f}$, $V_{3f}$, etc) are measured. The THD can then be calculated as:

\begin{equation}
THD(f) = {\sqrt{V_{2f}^2 + V_{3f}^2 + V_{4f}^2 + \cdots} \over V_f}\label{eq:THDDef}
\end{equation}
If the system contains only resistive components, the THD does not depend on the input frequency. For calculating the THD, we consider the range of load lineas as shown in Figure~\ref{f:methods}. From such a distribution, we estimate the average, $\uTHD$, and standard deviation, $\sTHD$, of THD.

Throughout the discussion, we also consider the average of the measurements for a given array. Such averages are indicated by a bar above the respective measurement. For instance, regarding the average, $\mu_{\beta}$, and standard deviation, $\sigma_{\beta}$, of individual transistors, the overall averages of these properties for all transistors in a given array are respectively represented as $\bar{\mu}_{\beta}$ and $\bar{\sigma}_{\beta}$.

Since we consider many properties to characterize the devices (i.e. the mean and standard deviation of $\beta$, $A$ and THD), it is useful to reduce the complexity of the data in order to aid the interpretation of the results. Such a task can be accomplished by using Principal Component Analysis (PCA)~\cite{joliffe1992principal}. Given a dataset containing $m$ features, PCA can be used to define a new, smaller, set of features $\mathrm{PCA}_1$, $\mathrm{PCA}_2$, $\dots$, $\mathrm{PCA}_k$ providing optimal preservation of the data variance. The first component ($\mathrm{PCA}_1$) contains the largest variance of the data, followed by $\mathrm{PCA}_2$ and so on. The amount of variance retained by each new feature can be calculated from the normalized $i$-th eigenvalue of the covariance matrix obtained from the data. This quantity, here represented as $E_i$ for $\mathrm{PCA}_i$, is defined in the range $[0\%,100\%]$.

While PCA implements a projection so as to optimize the data variability along the first axes, there are other projection schemes that optimize other properties.  Linear Discriminant Analysis~\cite{duda2012pattern} (LDA) is one of such schemes, which optimizes the linear separation between groups being, therefore (and unlike PCA), a \emph{supervised} method.

\section{Results and Discussion}

In this section, we present several analysis of the experimentally obtained characteristics of the arrays and respective BJTs.  We follow a progression from a simpler measurement-by-measurement characterization, to LDA, passing by PCA approaches.

\subsection{Measurement-by-Measurement Analysis}

\begin{figure*}[!htbp]
 \centering
 \subfigure[]{\includegraphics[width=0.48 \linewidth]{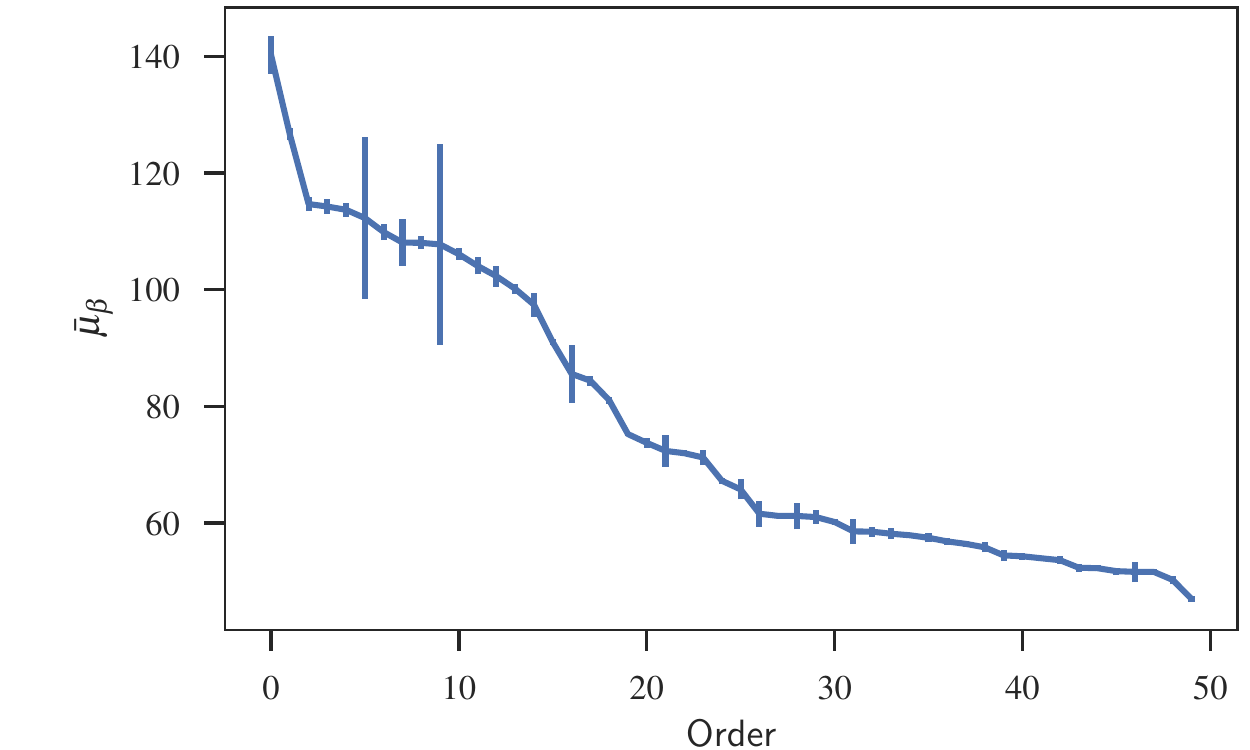}}~
 \subfigure[]{\includegraphics[width=0.48 \linewidth]{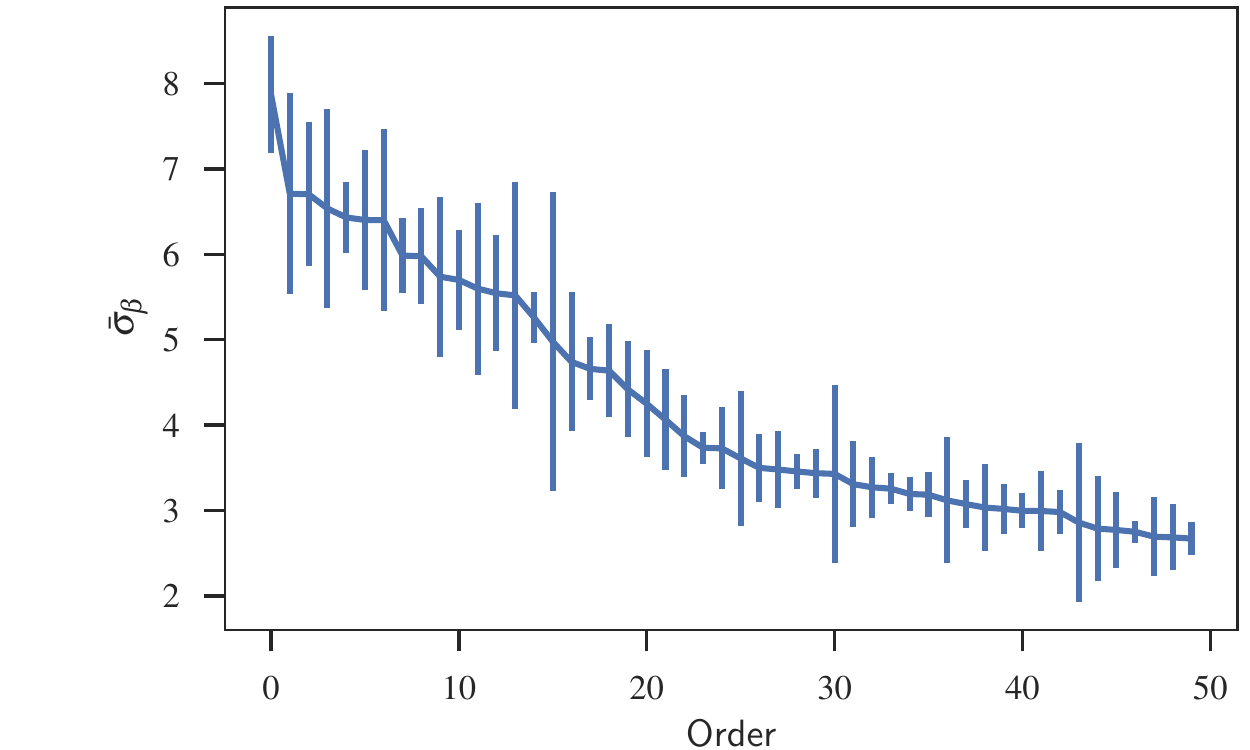}}\\
 \subfigure[]{\includegraphics[width=0.48 \linewidth]{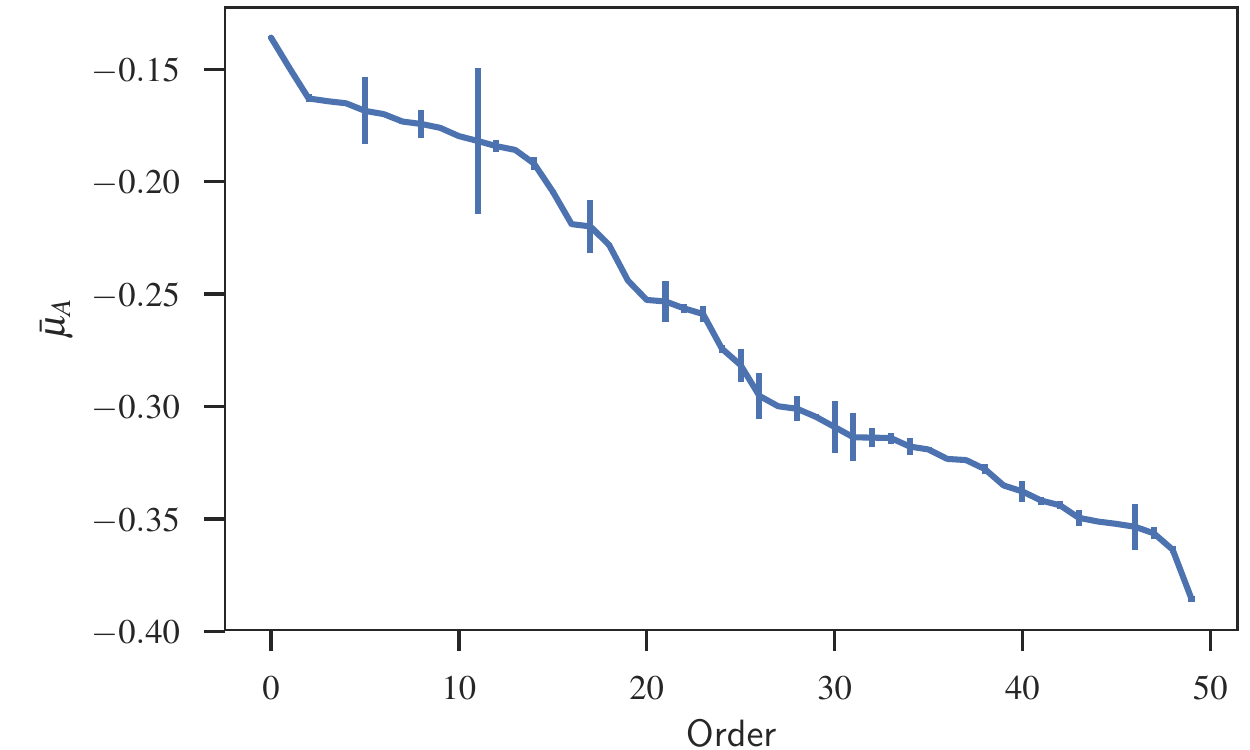}}~
 \subfigure[]{\includegraphics[width=0.48 \linewidth]{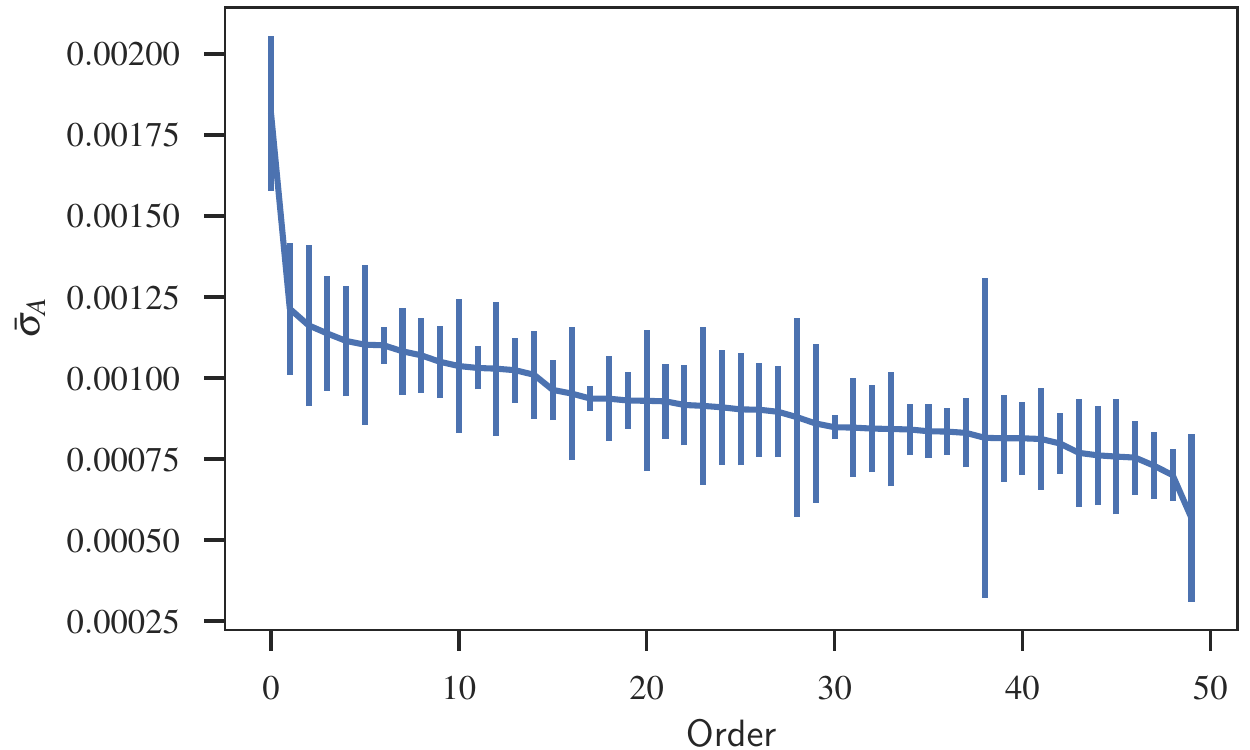}}\\
 \subfigure[]{\includegraphics[width=0.48 \linewidth]{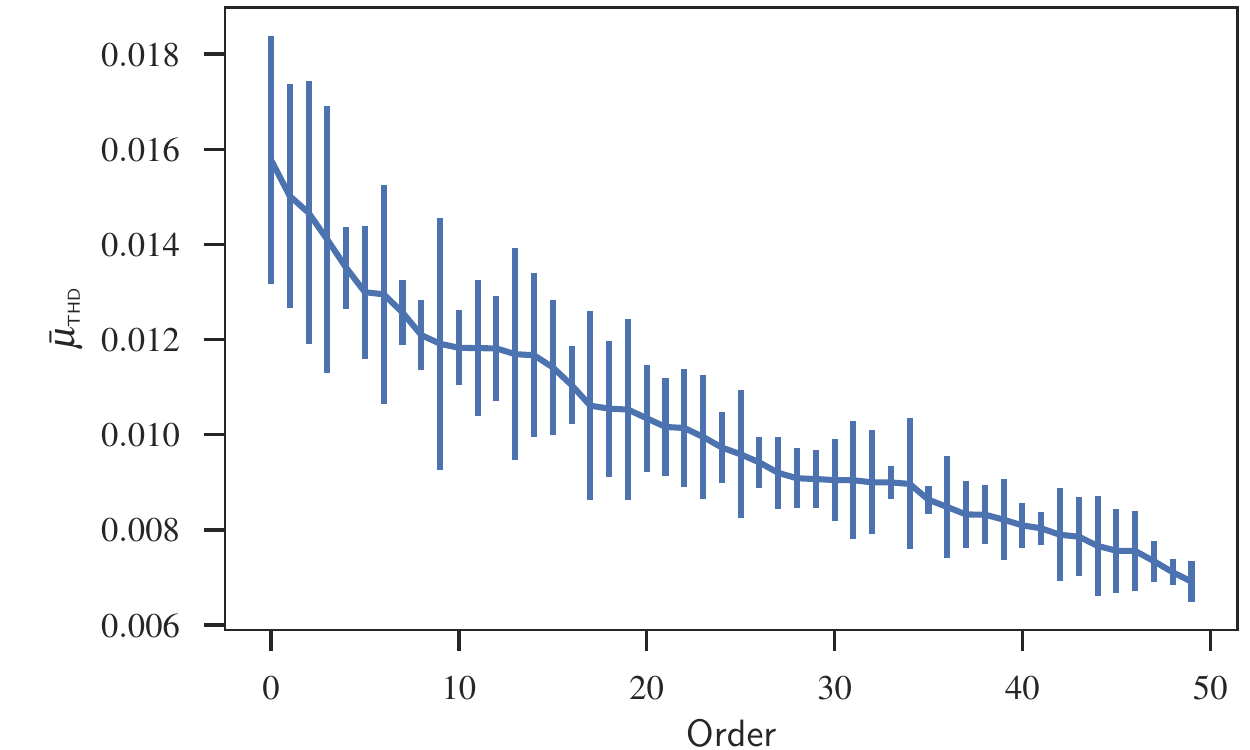}}~
 \subfigure[]{\includegraphics[width=0.48 \linewidth]{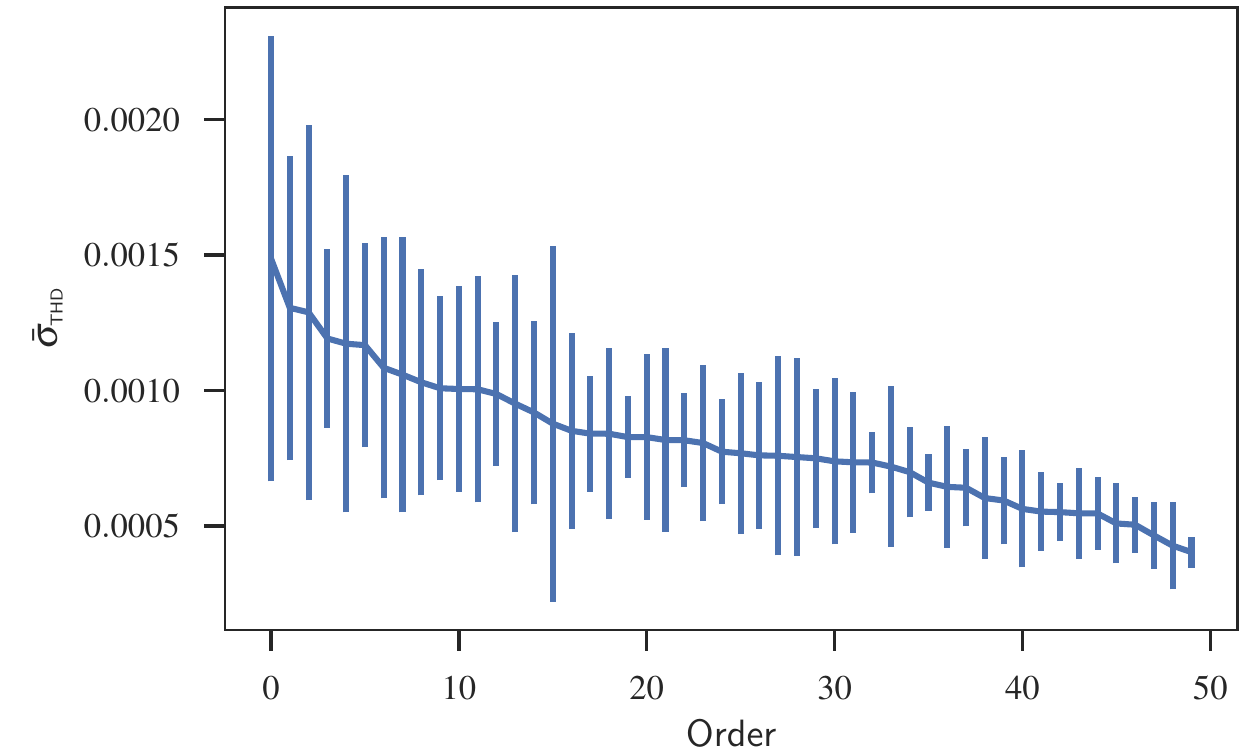}}
 \caption{The average (left-hand side) and standard deviation (right-hand side) of the current gain, voltage gain and THD obtained for the considered 300 BJTs.}
 \label{fig:avgstdarray}
\end{figure*}

Figure~\ref{fig:avgstdarray} shows the averages and standard deviations of the several considered measurements within each array. The x-axes are organized in monotonically decreasing order.  

Regarding the average current gain ($\mu_{\beta}$) values, shown in Figure~\ref{fig:avgstdarray}(a), we have that the group average range from 50 to 140, which are typical of BJTs.  Small variations of such averages are observed, except for a few cases.  The group average of $\sigma_{\beta}$, shown in Figure~\ref{fig:avgstdarray}(b), range from 3 to 8, which is very small considering the respective average values, suggesting that BJTs from a same chip are remarkably coherent regarding $\beta$.  The averages of the voltage gain properties, $\mu_A$ and $\sigma_A$, in common-emitter configuration are depicted in Figures~\ref{fig:avgstdarray}(c) and~\ref{fig:avgstdarray}(d).  The former varies from -0.40 to -0.15.  Figures~\ref{fig:avgstdarray}(e) and~\ref{fig:avgstdarray}(f) show the group average of $\uTHD$ and $\sTHD$. The averages range from 0.007 to 0.016, while the standard deviations go from 0.0005 to 0.0015.  All in all, these results suggest a remarkably small parameter variation within each array.

\begin{figure}
  \centering    \includegraphics[width=0.48\textwidth]{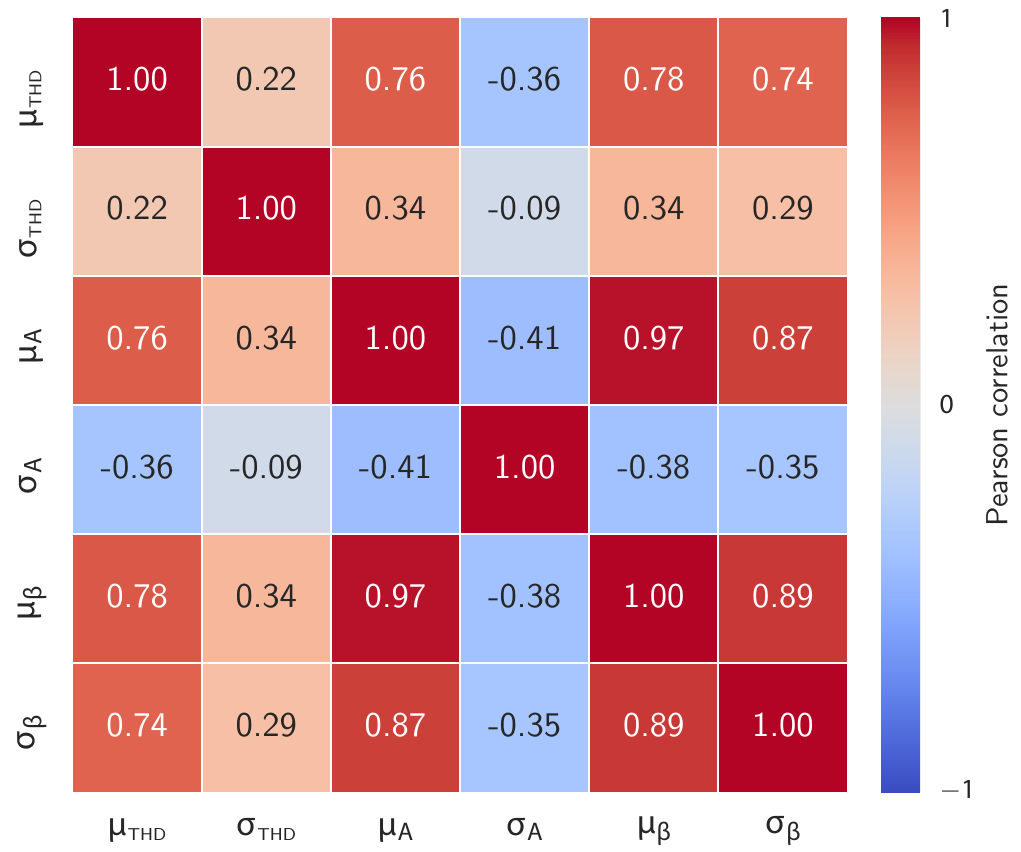}
  \caption{The Pearson correlation coefficients between the 6 considered BJT measurements.  Most measurements are correlated one another, except mainly for the standard deviation of the voltage gain}\label{f:correlations}
\end{figure}

Figure~\ref{f:correlations} shows the Pearson correlation coefficients between the several adopted measurements.   It is clear from this figure that most measurements are cross-correlated, except mainly for the standard deviation of THD and voltage gain, which are much more independent in this respect.  The most intense correlation results between the two gains, $\mu_{\beta}$ and $\mu_A$.  Observe that the voltage gain is often negatively correlated with the other measurements.

\subsection{PCA and Density}

\begin{figure*}
 \centering    \includegraphics[width=0.85\textwidth]{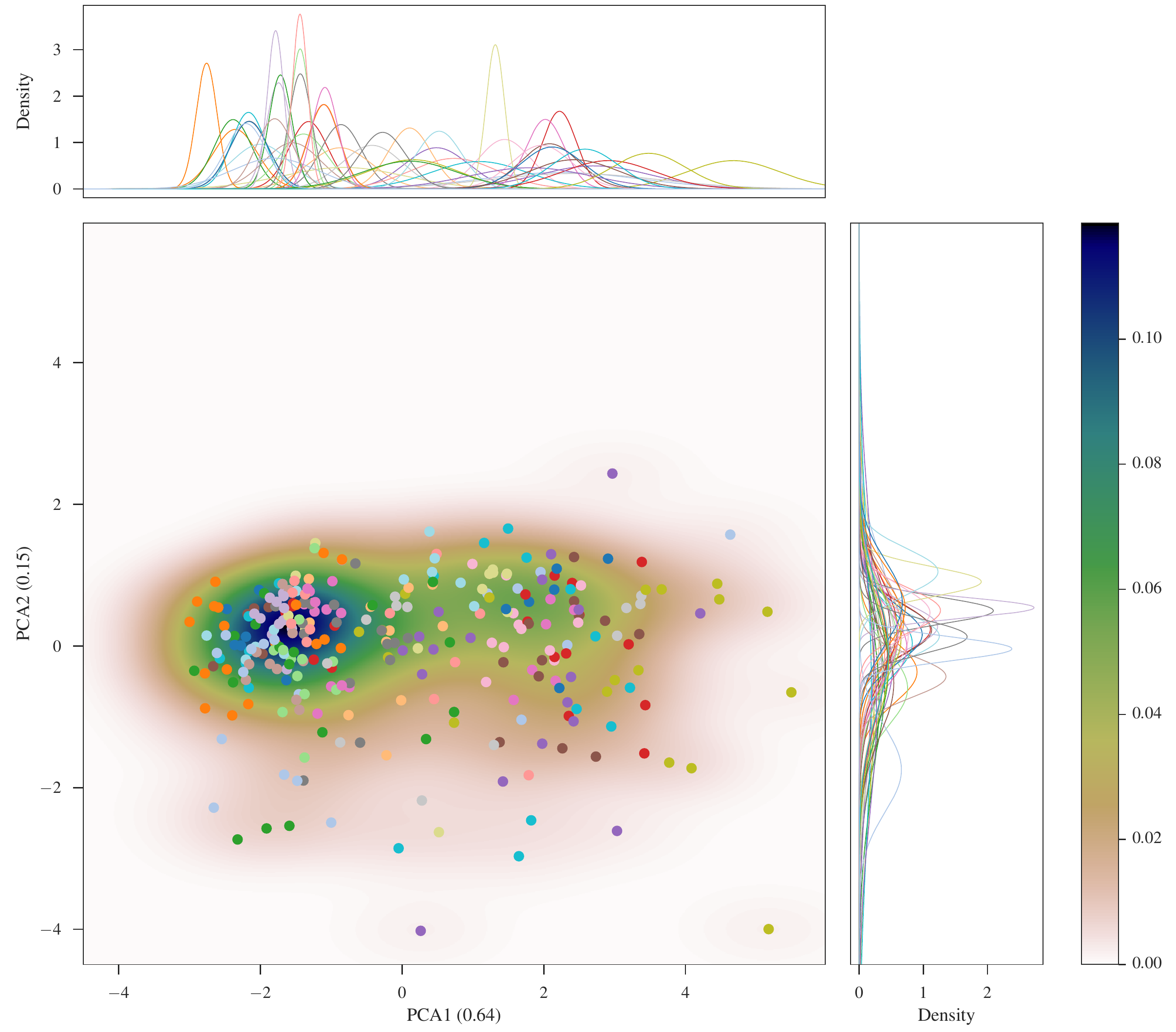}
 \caption{PCA obtained from the experimental measurements of the 50 samples of transistor arrays, as well as the individual clusters normally fitted and projected along both axes.  BJTs from a same array are shown by the same color.  The estimated density of cases is shown as a cloud, from which two peaks can be observed: a more compact group to the left-hand side, and a less compact to the right-hand side.}\label{f:pca}
\end{figure*}

In addition to understanding the variability, in absolute manner, of the several measurements, it is also interesting to infer what is the effective dimensionality of the overall variability of the measurements after completely eliminating the correlation between them.  This can be readily achieved by using PCA.  Figure~\ref{f:pca} shows the PCA projection of the 300 BJT samples considered in our experiments.

Several interesting findings can be derived from such a result. First, the total variation explanation for the two first axes, respectively equal to $64\%$ and $15\%$, indicate that the greatest part of the variation of the transistor parameters can be accounted for by nearly just two degrees of freedom, corresponding to respective random variables defined by linear combinations of the adopted measurements.  This result is in agreement with previous experiments~\cite{costa2016negative} suggesting that BJT variability has a low dimensionality.

Another important result concerns the fact the BJTs belonging to the same array tend to appear near one another, defining respective groups in the PCA space.  As shown in the projections respective to each of the PCA axes, the separation between arrays is much more definite along the first axis than for the second.  These projections were obtained by normally individually fitting each group (corresponding to each transistor array) and then mapping into the first and second PCA axes.  The dispersion of each group, as indicated in the first axis projection, tends to increase along that axis.

It is also clear that the parameter variance within each IC is substantially smaller than the variance considering all devices.  Henceforth, we quantify the variance of a group of individuals by taking the trace of the respective covariance matrix, giving rise to the \emph{total variation} measurement.  A measurement of how effectively the parameter variation can be minimized by adopting transistors from a same IC can therefore be expressed in terms of the ratio between the total variation for all devices divided by the total variation averaged among the IC clusters.  This value was obtained as being nearly 5 for the first PCA component.  

It is also clear from Figure~\ref{f:pca} that the overall distribution of BJTs is skewed, with a stronger peak at the left-hand side and a secondary, less compact, peak to the right.  The first peak is characterized by $\text{PCA1} \approx -1.8$, and the second by $\text{PCA1} \approx 2.1$.

\begin{figure*}
 \centering    \includegraphics[width=0.85\textwidth]{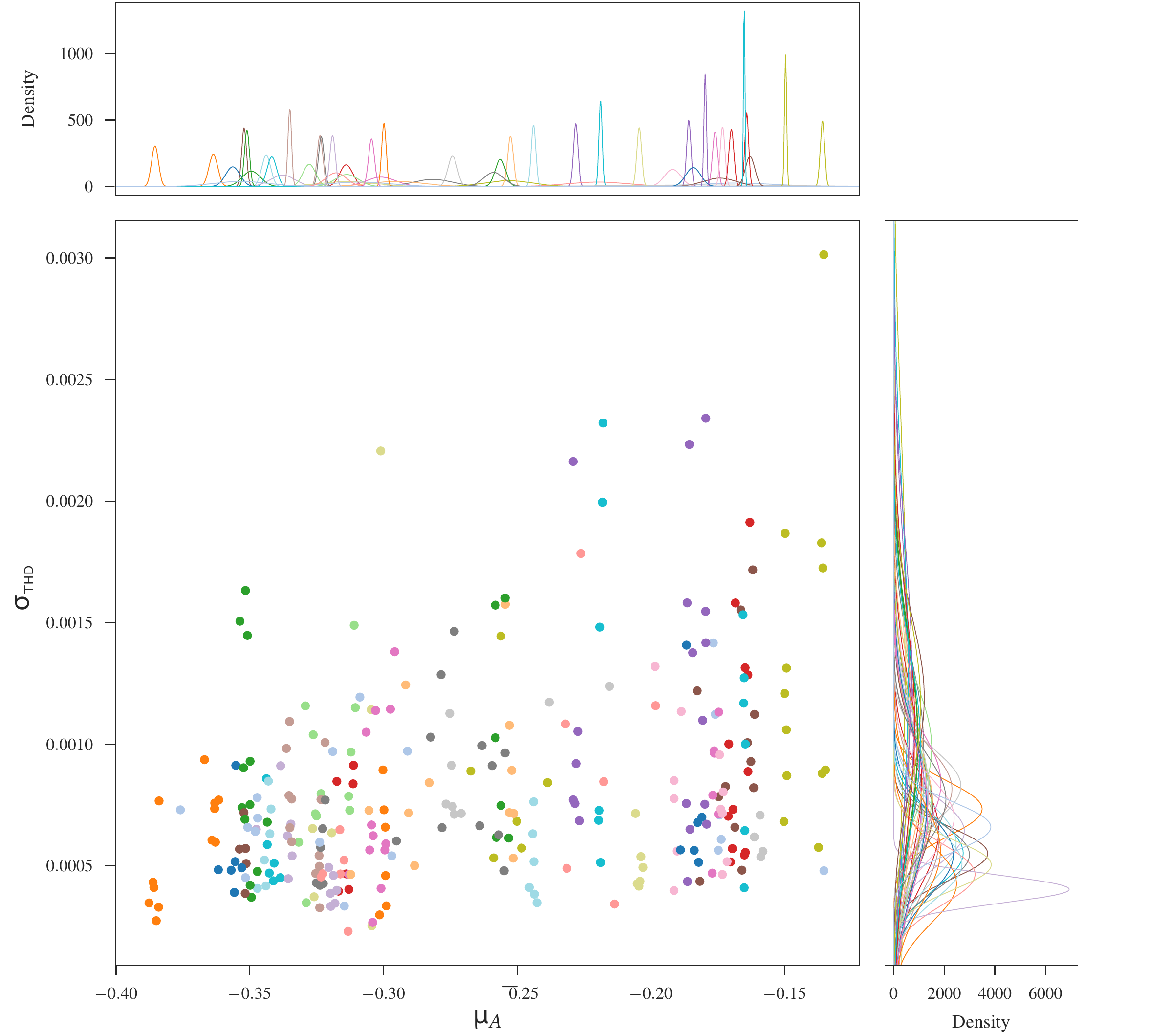}
 \caption{Scatterplot obtained by plotting the variation of the THD in terms of the average of the voltage gain, also showing normally fitted individual clusters along both axes.  Such a mapping, defined by two measurements suggested by the PCA, yielded a remarkable separation between the transistor arrays, especially along the average voltage gain axis.  The average variation of such individual clusters was verified to be nearly 20 times smaller than the overall BJT variance considering all the clusters.}\label{f:gainthd}
\end{figure*}

The weights of each considered measurement on each axis are given in Table~\ref{t:weights}.  The first axis is strongly related to the average THD, voltage gain, and current gain, as well as the standard deviation of current gain.  Actually, we have from Figure~\ref{f:correlations} that these four variables are strongly correlated.  In the case of the second axis, the major contribution is from the standard deviation of THD.  Such results suggest to use the average voltage gain and standard deviation of THD in order to summarize the first and second axes, respectively.   The respectively obtained scatter plot is shown in Figure~\ref{f:gainthd}.   Remarkably, the arrays resulted very compact and segregated along the average voltage gain, with a gain factor of about 20 in parameter variation along the gain axis (observe that the current and voltage gain resulted highly correlated).

\subsection{LDA}

\begin{figure*}
 \centering
 \includegraphics[width=0.85\textwidth]{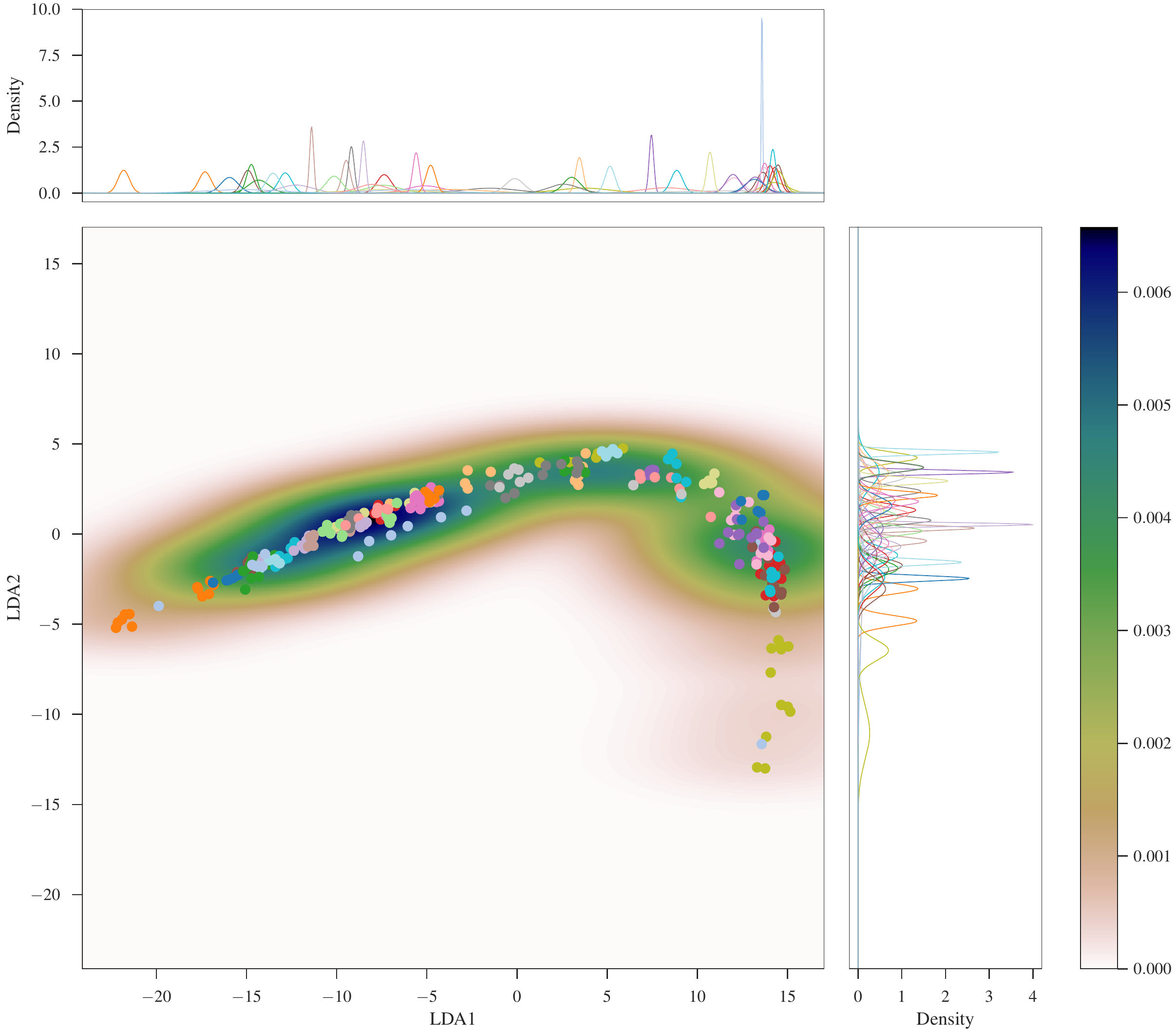}
 \caption{LDA of the 300 BJTs, as well as normally fitted individual clusters along each axis.  A good separation of the groups is obtained, and the mapping also revealed a clear functional relationship between the two respective axes.}\label{f:lda}
\end{figure*}

\begin{figure}
 \centering    \includegraphics[width=0.45\textwidth]{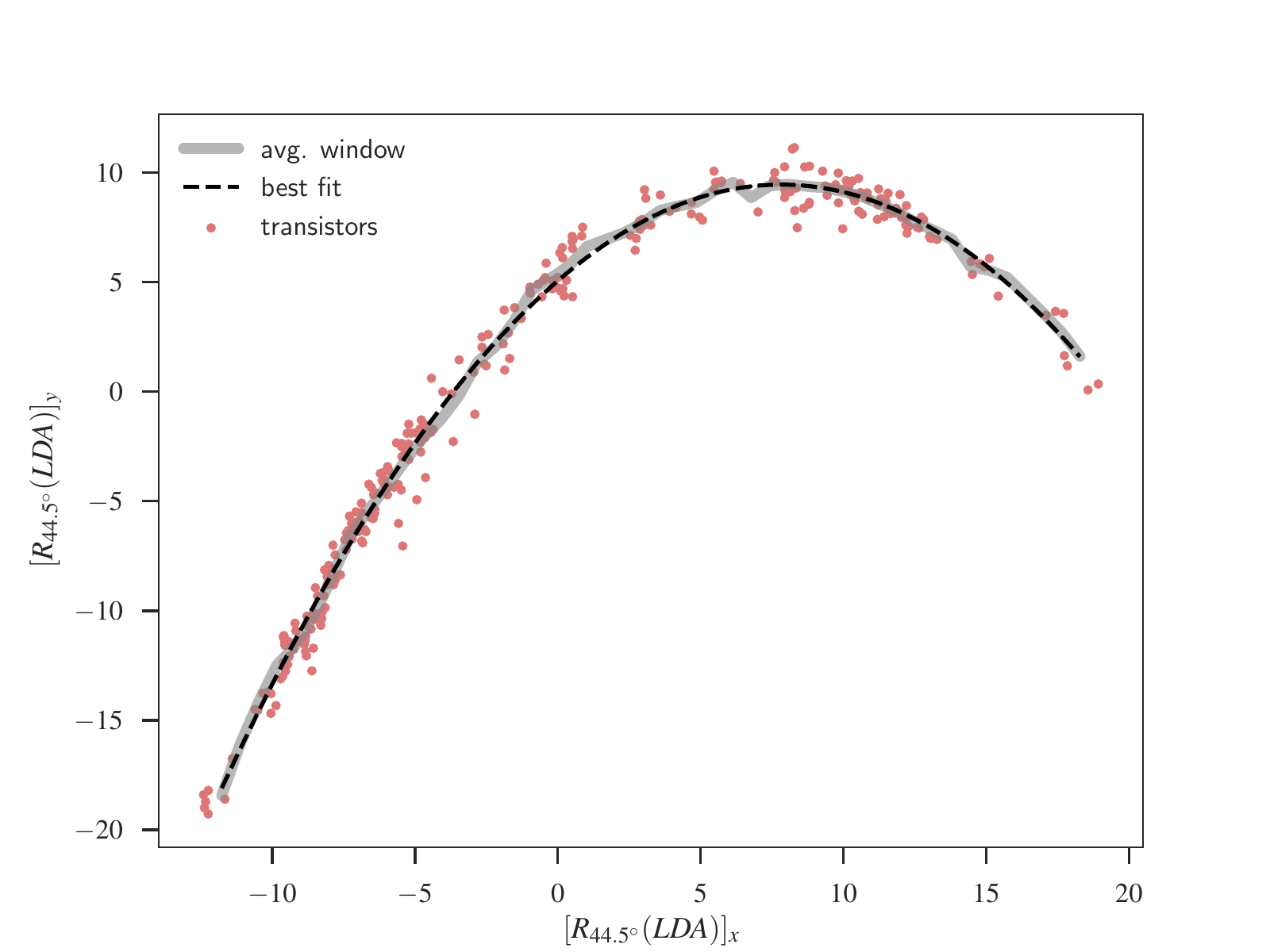}
 \caption{Generalized (quadratic) least mean square fitting of the LDA results, after rotation optimizing the fitting residue.}\label{f:ldaBestFit}
\end{figure}

The LDA analysis of the 300 BJTs yielded the two-dimensional projection shown in Figure~\ref{f:lda}.   At the expense of a less effective minimization of variance along the first two axes, LDA provided a new projection that maximizes the separation of groups according to scatter matrices dispersions.  At the same time, the dispersion within each group became much more uniform and constrained, to the point of defining a narrow relationship between the two axes, which can be summarized by the dashed line.  In order to try obtaining a functional relationship between the adopted measurements, we applied generalized mean least square fitting~\cite{draper1966applied,bjorck1996numerical} while rotating the two-dimensional LDA projection and seeking for the smallest fitting residual. For this, we used a second degree polynomial as the template for the fitting. In order to avoid the effect of the varying density of samples in the LDA space, windowed weighting of one axis in terms of the other was applied for each rotating configuration. Figure~\ref{f:ldaBestFit} shows the obtained optimal fitting. 

The obtained fitting coefficients for the first two LDA components resulted in the following approximate relationship between LDA1 and LDA2:
\begin{multline}
\text{LDA1}^2 +  \text{LDA1} (-2.723 - 1.961\times\text{LDA2}) \\+ (40.93 + 0.9612\times\text{LDA2})\times\text{LDA2} = 138.2, \label{eq:LDAFitting}
\end{multline}
which can also be written approximately in terms of the considered transistor parameters as
\begin{multline}
11 \mu _A + \mu _A^2 -0.112 \mu _A \mu _{\text{\tiny THD}}-13 \mu _{\beta } \\
+1.85 \mu _A \mu _{\beta }-0.104 \mu _{\text{\tiny THD}} \mu _{\beta } +0.857 \mu _{\beta }^2 \\ + 0.111 \mu _A \sigma _A +0.103 \mu _{\beta } \sigma _A-0.177 \sigma _{\text{\tiny THD}} \\ = 3.516
\label{eq:LDAFitProps}
\end{multline}

\begin{table}
	\caption{The weights (or strengths) of each of the considered measurements on the first and second axes of the PCA and LDA.  The larger the magnitude of such weights, the larger its effect  in defining the respective axis, which are linear combinations of the original measurements.}
	\label{t:weights}
	\centering
	\begin{tabularx}{\columnwidth}{XXXXXXX}
		\hline
		\textbf{Axis} & $\bm{\mu_{\beta}}$ & $\bm{\sigma_{\beta}}$ & $\bm{\mu_{A}}$ & $\bm{\sigma_{A}}$ & $\bm{\uTHD}$ & $\bm{\sTHD}$ \\
		\hline
		PCA1 & 0.49 & 0.47 & 0.49 & -0.26 & 0.44 & 0.21 \\
        PCA2 & 0.01 & -0.01 & 0.00 & 0.52 & -0.11 & 0.85 \\
        LDA1 & -6.86 & 0.19 & 17.81 & 0.22 & 0.03 & 0.00 \\
        LDA2 & -12.92 & 0.1 & 11.77 & -0.13 & 0.39 & -0.17 \\
		\hline
	\end{tabularx}
\end{table}
   
\section{Conclusions}

Although at the core of modern electronics, the issue of transistor variability has received relatively little attention as a consequence of the effectiveness of negative feedback in controlling such an effect.  However, recent studies~\cite{costa2016negative} have suggested revisions of such a perspective in the sense that BJTs of a given type can have relatively consistent features (small variations), capable of defining clusters by transistor type.  In other words, BJTs from the type can cluster in characteristic, distinctive groups.  The current work has investigated BJT variability further, now considering transistor arrays, where the devices share the substrate and other fabrication effects.  As it could be expected, the variability of BJTs inside the same IC was found to be much smaller (by a factor of 20 in the case of current gain, for the considered devices and configuration) than between BJTs from different arrays.  This confirms the improved control of transistor properties when built into the same IC.  At the same time, larger variations were observed among BJTs from different arrays. 

It should be emphasized that, though in principle expected, the reduced variability of industrially produced BJTs inside a same chip has been rarely (if ever) systematically quantified in the literature.  The obtained results make it clear, at least for the considered array samples, that a substantial gain in standardization of BJT parameters can be achieved by using transistor arrays.  However, critical projects large enough to involve multiple arrays may require the consideration of the individual characteristics of the used devices, as a consequence of the larger variation of parameters among BJTs from distinct arrays.   The obtained results corroborated the potential of integrated analog design for increasing parameter uniformity.  In addition, when the data was mapped in terms of the scatterplot defined by the THD variation and the voltage gain (as suggested by the strength of these measurements respectively to the two PCA axes), an even better separation was obtained between the transistor arrays, showing most of the clusters clearly seggregated one another along the latter axis, while similar distributions were observe in the former axis.  

Further analyses by using LDA showed an even better separation between the clusters corresponding to respective transistor arrays.   Interestingly, the LDA mapping revealed a clearly defined relationship between the considered measurements, probably related to the Early effect~\cite{early1952effects,boylestad2002electronic}.  This result, as well as all the others in the current work, are respective to the considered transistor arrays samples and circuit configuration, so that further investigations are required in order to validate and further generalize them.  

Perhaps as a consequence of well-established design practices, some important issues in analog electronics have been somewhat overlooked.  It is, therefore, interesting to revisit such issues in the light of state-of-the-art pattern recognition and statistical analysis concepts and methods.  It would be particularly interesting, for instance, to complement the currently reported study by considering other IC families (e.g. FET, MOSFET) and alternative circuit configurations, as well as to compare parameter variability between NPN and PNP devices.

\section*{Acknowledgements}
L. da F. Costa thanks CNPq (Grant no. 307333/2013-2) for support. F. N. Silva acknowledges FAPESP (Grant No. 15/08003-4). C. H. Comin thanks FAPESP (Grant No. 15/18942-8) for financial support. This work has been supported also by FAPESP grant 11/50761-2.

\bibliography{referencias}

\end{document}